\documentclass[preprint,showpacs,preprintnumbers,amsmath,amssymb]{revtex4}
\usepackage{epsfig}

\begin{document}
\draft
\title{The Inductive Single-Electron Transistor (L-SET)}
\author{Mika A. Sillanp\"a\"a}
\author{Leif Roschier}
\author{Pertti J. Hakonen}
\affiliation{Low Temperature Laboratory, Helsinki University of
Technology, Otakaari 3 A, Espoo P.O.Box 2200 FIN-02015 HUT
Finland}

\begin{abstract}

We demonstrate a sensitive method of charge detection based on
radio-frequency readout of the Josephson inductance of a
superconducting single-electron transistor. Charge sensitivity
$1.4 \times 10^{-4}$e$/\sqrt{\mathrm{Hz}}$, limited by
preamplifier, is achieved in an operation mode which takes
advantage of the nonlinearity of the Josephson potential. Owing to
reactive readout, our setup has more than two orders of magnitude
lower dissipation than the existing method of radio-frequency
electrometry. With an optimized sample, we expect uncoupled energy
sensitivity below $\hbar$ in the same experimental scheme.
\end{abstract}


\maketitle

Quantum measurement in the solid state has been shown to be
feasible as several realizations of qubits based on mesosocopic
superconducting tunnel junctions have emerged
\cite{nakamuraqb,vion,hanqb,martinisqb}. Measurement of physical
quantities close to the limit set by the uncertainty principle is,
on the other hand, an important issue in its own right. The
detector should not only have a high gain, but also its internal
noise should not act back on the measured observable more than
allowed by the uncertainty principle.

The Single-Electron Transistor (SET) is a basic mesoscopic
detector, sensitive to electric charge on a gate capacitor. Its
operation is based on stochastic tunneling of single electrons,
typically at sub-Kelvin temperatures $k_B T \ll e^2 / (2
C_{\Sigma})$, where $e$ is the electron charge and $C_{\Sigma}$
total capacitance of the SET. Due to shot noise the SET has,
however, a quite substantial back-action, and even theoretically,
its energy sensitivity \cite{ds} remains approximately a factor of
four from the fundamental quantum limit $\hbar / 2$.

The SET has a practical problem with a low bandwidth, but a few
years ago it was demonstrated \cite{rfset} that a band of tens of
MHz is possible with radio-frequency SET (rf-SET) where impedance
match from a high-impedance SET to $50 \, \Omega$ is achieved by
means of an $LC$ tank circuit. Several picowatts of power is
dissipated in the rf-SET under optimal working conditions, which
warm up the SET island up to half a Kelvin.

Cooper pair tunneling, in contrast, is correlated, and hence a
detector based solely on the Josephson effect does not exhibit
shot noise or dissipation. Zorin has introduced a superconducting
quantum-limited low-frequency electrometer \cite{zorin,zorinexp},
and a theory for a corresponding high-frequency device
\cite{zorinrf} where readout is performed similarly as in the
rf-SQUID (see also \cite{cottet}). Any experimental
demonstrations, nevertheless, of non-dissipative high-frequency
electrometry have been lacking to date.

In this paper we present a new type of such device, the Inductive
Single-Electron Transistor (L-SET), and demonstrate it in
experiment. Two distinct operation modes are identified. The
"plasma oscillation" mode corresponds to small oscillations of the
Josephson phase $\varphi$ in harmonic potential, whereas in
"non-harmonic" mode, higher excitation is used such that the swing
is more than $2 \pi$. Charge sensitivities of $2.0 \times
10^{-3}$e$/\sqrt{\mathrm{Hz}}$ and $1.4 \times
10^{-4}$e$/\sqrt{\mathrm{Hz}}$ were achieved in the two modes,
respectively. The present experiment is also the first one to
directly probe the nonlinear dynamics of a mesoscopic
superconducting junction, in terms of both the amplitude and phase
of microwave voltage reflection coefficient $\Gamma$.

\begin{figure}
  \includegraphics[width=7cm]{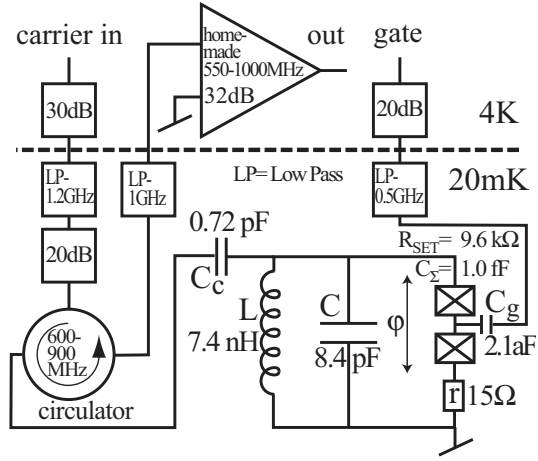}
  \caption{Schematics of the L-SET circuitry. A SSET is coupled in
  parallel with an $LC$ circuit resonant at $f_0 = 613$ MHz.}\label{fig:circuit}
\end{figure}

A superconducting SET (SSET, also called Cooper-Pair Transistor)
has two small-capacitance Josephson junctions in series, and the
Hamiltonian is $(q-q_g)^2 / (2 C_{\Sigma}) - 2 E_J \cos(\varphi/2)
\cos(\theta)$. Here, $q_g$ is the gate charge, $\varphi$ is the
phase difference across the whole device, assumed to be a
classical variable due to an environment having impedance much
smaller than $R_Q = h/(2 e)^2 \simeq 6.5$ k$\Omega$, whereas
$\theta$, conjugate to the charge of the island $q$, experiences
quantum effects. Evidence of the resulting band structure $E_n
(\varphi, q_g)$ has been obtained in experiment \cite{lukens}. The
lowest band $n = 0$ has approximately sinusoidal energy $E_0 (
\varphi, q_g)$, and thus, the SSET is effectively a gate-tunable
single junction \cite{brink}. From our point of view the most
important property of SSET is its Josephson inductance $L_{J}
^{-1} (\varphi, q_g) = \left( 2 \pi / \Phi_0 \right) ^2 \partial^2
E (\varphi, q_g) /
\partial \varphi^2$, where $\Phi_0 = h/(2 e)$ is the flux quantum.
With a total shunting capacitance $C$, a SSET thus forms a
harmonic oscillator with the plasma frequency $f_p = 1/(2 \pi)
(L_J C)^{-1/2}$ for small $\varphi$.

In our experiment, $f_p$ is tuned into an experimentally
accessible range, below one GHz, by a large $C \simeq 8.4$ pF. The
SSET is also shunted with an external inductor $L \simeq 7.4$ nH
which offers several theoretical as well as practical advantages.
The circuit (see Fig.\ \ref{fig:circuit}) is coupled to feedline
via a coupling capacitor $C_c \ll C$. The sources of dissipation
are lumped into the resistor $r$ in series with the SSET.

Energy of the system of the SSET plus the $LC$ oscillator is the
sum of potential energy $E_{n} (\varphi, q_g)  + \Phi^2 / (2 L)$,
approximately $- E_{0} (q_g) \cos(\varphi) + \left[\Phi_0 ^2 /(4
\pi^2 L)\right] \varphi^2$ at the ground band, and kinetic energy
$q^2 / 2 C_{\Sigma}$. Here, $\Phi$ is the magnetic flux in the
loop. Assuming the SSET stays at the lowest band, where $I \simeq
I_0 \sin (\varphi)$, classical dynamics of phase in the oscillator
is thus analogous to a particle moving in a sinusoidally modulated
parabolic potential. The dynamics is similar to that in the
rf-SQUID \cite{erne}, i.e., a single junction shunted with a loop
inductance, except that since our loop is not fully
superconducting, flux quantization or flux jumps do not exist. At
small driving amplitude (linear regime), the phase particle
experiences harmonic oscillations around $\varphi = 0$ at the
frequency $f_p = 1/(2 \pi) (L
\parallel L_{J} C)^{-1/2}$ which is controlled by gate-tuning
$L_J$. This mode of operation, where the L-SET works as a
charge-to-frequency transducer, we call the "plasma oscillation
mode"

At higher oscillation amplitude, the particle sees a different
local curvature of the potential due to the $\cos(\varphi)$ term,
thus changing the oscillation frequency. At very high amplitude,
the cosine modulation becomes effectively averaged out. Thus, when
increasing excitation, we expect a change of resonant frequency
from $f_p$ to $f_0 = 1 /(2 \pi) (L C)^{-1/2}$. This change of
resonance frequency at a critical excitation power $P_c$ can be
seen as an analogue of a dc-biased Josephson junction switching
into a voltage state. Qualitatively, this type of behavior was
identified in the experiment.

At large excitations, the highly nonlinear oscillator experiences
complicated dynamics which do not in general allow analytical
solutions. Numerical calculations over a large range of
parameters, however, show consistently that the system response
depends on $L_J$ also in this case \cite{tbb}. This is what we
call the "non-harmonic" mode.

\begin{figure}
  \includegraphics[width=7cm]{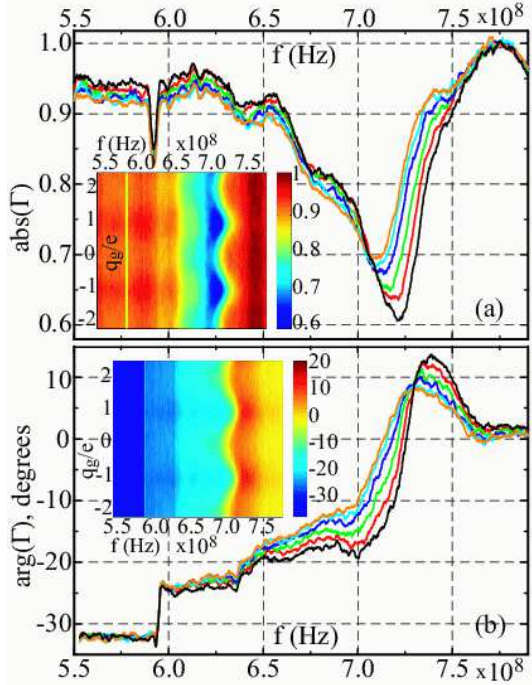}
  \caption{Measured frequency response at $-125$ dBm for successive gate
  values $q_g/e = 1 \ldots 0$ from bottom to top, (a) amplitude, (b)
  phase. Effective resolution bandwidth $\simeq 1$ Hz.
  The insets portray the periodicity.}\label{fig:response}
\end{figure}

The Al SSET was of a standard design, incorporating Cu
quasiparticle traps 4 $\mu$m from the junctions, positioned such
that they are not on the conduction path. From the total tunnel
resistance $R_{SET} = 9.6 $ k$\Omega$ we get the single-junction
Josephson energy $E_J = 1.56$ K. Surface area of the two tunnel
junctions $0.021 \, \mu$m$^2$ gives an estimate of the charging
energy of the SET (capacitance $\sim 43$ fF$ / \mu$m$^2$), $E_C
\approx 1.0$ K. However, this method does not allow sufficient
accuracy due to substantial sensitivity of the electrometer
performance to $E_J / E_C$. The best fit was obtained with $E_J /
E_C = 1.7$, thus $E_C = 0.92$ K and $C_{\Sigma} = 1.0$ fF, which
is within error margins of the previous method. Value of each tank
circuit component was determined prior to cooldown to roughly 20
\% accuracy. These agreed with values got by fitting to frequency
responses.

The measurements were done in a dilution refrigerator at a base
temperature of 20 mK. Every effort was made to eliminate noise.
The microwave feedline was heavily attenuated and low-pass
filtered. A circulator at 20 mK, tested to have more than 20 dB
backward isolation, was used to cut back-action noise from the
cold amplifier. The coaxial gate lead was filtered and attenuated
at 4 K, and it had small capacitance $C_g \simeq 2.1$ aF in order
to avoid voltage noise.

We used microwave reflection to read the reactance of the
oscillator. $\Gamma = (Z - Z_0)/(Z + Z_0)$, where $Z_0 = 50 \,
\Omega$ and $Z$ is the sample impedance, was measured by probing
the system with a carrier wave, typically $\leq - 115$ dBm. The
reflected carrier was amplified with a chain of amplifiers having
a total of 5 K noise temperature, and detected with a network
analyzer or a spectrum analyzer.

At low excitation, we measured the plasma resonance at $f_p = 723$
MHz, and its shift with gate of 15 MHz (see Fig.\
\ref{fig:response}). With the best-fit tank circuit parameters,
the shift corresponds to a 15 \% modulation of $L_J$, achieved
according to theory at $E_J / E_C = 1.7$, in good agreement with
independently determined SSET parameters. For this $E_J / E_C$ we
calculate the minimum (with respect to $q_g$) value of the SSET
inductance $L_{J} = 15.4$ nH. Placed in parallel with $L$, we thus
expect $f_p = 744$ MHz, which agrees rather well with the
experimental value. The small disagreement can be due to
uncertainty in the values of $L$ and $C$.

The frequency response was periodic with respect to gate voltage
with a perfect $2e$ period. By increasing temperature, we observed
a transition to full $e$-periodicity at 300 mK as usually in a
SSET. $Q_e$, determined as the half width of the resonance,
reduced from 18 to 12 \cite{qnote} while tuning by gate from the
maximum to minimum in $f_p$, which is presently not understood.

\begin{figure}
  \includegraphics[width=7.5cm]{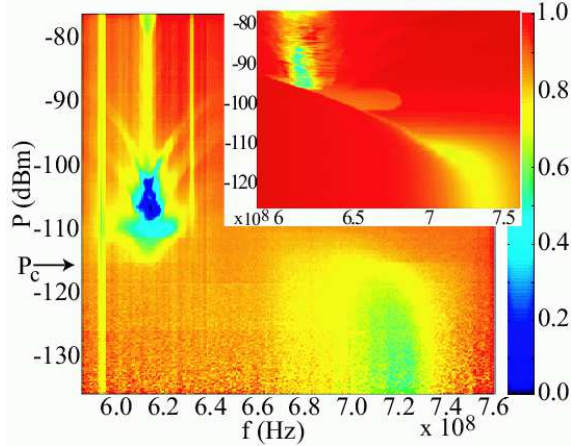}
  \caption{Measured frequency response ($ | \Gamma | $) as a function of excitation
  power. "Critical power" $P_c$ (see text) is marked.
  The inset shows a corresponding simulation made
  using the circuit of Fig.\ \ref{fig:circuit} (to aid convergence, we
  used a $2$ k$\Omega$ resistor across the SSET, and 10 k$\Omega$
  across the resonator).}\label{fig:freqpwr}
\end{figure}

Changes in frequency response (see Fig.\ \ref{fig:freqpwr}) as the
excitation is increased are in qualitative agreement with
classical dynamics. The wavelike texture at $-105 \ldots -90$ dBm
is due to the $\cos ( \varphi )$ Josephson-term. Changes in
coupling, e.g., that $Z$ goes through critical coupling at $-105$
dBm, and sharpening of the resonance are related to an increase of
$Q_i$ from 20 below $P_c$, up to several hundreds. The increase
could be reproduced in simulations involving only nonlinear
dynamics, but, only up to a factor of five approximately. We
anticipate some complex processes involving higher bands of the
SSET may be involved. The abrupt switch of resonance frequency at
$P_c$ in experiment is explained as being due to the same effect
(see later). A numerical harmonic balance simulation run using the
\textsc{Aplac} circuit simulator is shown as inset in Fig.\
\ref{fig:freqpwr}. The satellite dips $\pm 19$ MHz around the tank
circuit resonance arise probably from a spurious low-frequency
resonance that couples to the system.

The only internal source of noise and back-action are fluctuations
in the resistor. At 20 mK these are already at the quantum limit
$\hbar \omega \approx k_B T$. Although the cold amplifier has a
noise temperature $T_N \gg T$, temperature in the sample is only
approximately doubled to 40 mK by leakage through the isolator.
Noise in the output, however, is fully dominated by the
amplifiers. This is converted into an equivalent charge
sensitivity $s_q$ or energy sensitivity $\epsilon = s_q ^2 / (2
C_{\Sigma})$ by $s_q = s_V / (\partial V /
\partial q)$ where $s_V = \sqrt{Z_0 k_B T_N}$ in amplitude readout
of the reflected carrier voltage $V = V_0 | \Gamma |$. In readout
of the phase $\phi = \arg (\Gamma)$ we have $s_q = s_{\phi} /
(\partial \phi / \partial q)$, where $s_{\phi} = \sqrt{2} s_V /
V$.

\begin{figure}
  \includegraphics[width=7cm]{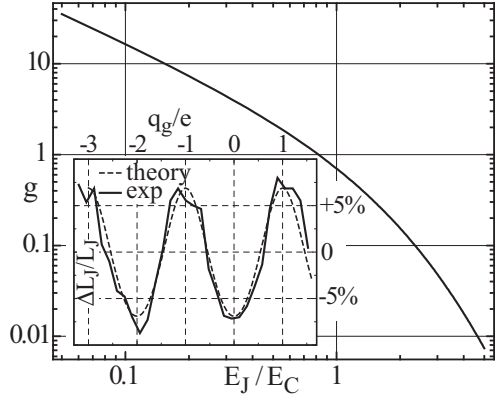}
  \caption{Electrometer gain $g$ calculated as a
  function of $E_J / E_C$. The inset shows gate
modulation of $L_J$ for the present sample.}\label{fig:gain}
\end{figure}

The derivatives can be written as $\partial V / \partial q = V_0
(\partial | \Gamma | / \partial l_{J}) (\partial l_{J} / \partial
q)$, where $l_{J} = L_J / \min(L_{J})$, and similarly for the
phase. In the plasma oscillation mode, $\partial | \Gamma | /
\partial l_{J}$ is computed by standard circuit analysis, but in
the non-harmonic mode, we rely on numerical simulations. The
"gain" $g = \max \left( \partial l_{J} / \partial q \right)_{q_g}
e$, expressed as relative modulation of the Josephson inductance
per electron, is the main figure of merit of the SSET
electrometer, $g = 0.23$ for the present sample according to the
theory (Fig.\ \ref{fig:gain}).

In the plasma oscillation mode, the optimal excitation ($P_c$)
corresponds to critical ac-current. For the present sample we
calculate in detail \cite{tbb}, both in amplitude or phase
readout, $s_q = 5 \times 10^{-4}$e$/\sqrt{\mathrm{Hz}}$. In the
non-harmonic mode, $\partial | \Gamma | / \partial l_{J}$ is
smaller according to simulations, but the sensitivity is predicted
to be better by an order of magnitude, $s_q \approx 4 \times
10^{-5}$e$/\sqrt{\mathrm{Hz}}$, due to the possibility for a high
excitation.

Charge sensitivity was measured in a straightforward manner using
amplitude readout. A small $1/20 e_{RMS}$ marker signal was fed to
the gate at a frequency $f_g$. The modulated carrier was detected
with spectrum analyzer, and signal-to-noise ratio was read from
side bands at $f_p \pm f_g$. In the plasma oscillation mode we got
$s_q = 2.0 \times 10^{-3}$e$/\sqrt{\mathrm{Hz}}$  at the maximum
power $P_c \simeq -116$ dBm, corresponding to 20 fW dissipation in
the whole resonator circuit. Note that due to Cooper pair
tunneling, the power is not dissipated in the SSET island. In the
non-harmonic mode, significantly better sensitivities were
obtained. $s_q = 1.4 \times 10^{-4}$e$/\sqrt{\mathrm{Hz}}$ was
measured at $-100$ dBm \cite{sample}. Due to a higher impedance
mismatch, power dissipation can be kept below 20 fW in this case
as well.

The sensitivities were measured at $f_g = 300$ kHz, however, 3 dB
bandwidths $\Delta f$ were roughly 40 MHz in plasma oscillation
mode and 10 MHz in non-harmonic mode, in agreement with inverse
response time of the oscillator, $\Delta f \simeq f_0 / Q_e$.

The discrepancy of a factor of 4 in the predicted sensitivities is
partially explained through another process not included in the
model. Namely, according to the simple model, inverse sensitivity
should increase linearly with amplitude. A clearly weaker
dependence was found in the experiment, however. At small
amplitude, theory and experiment are within a factor of 2 from
each other. We attribute the effect to an increased rate of
Zener-transitions to a higher band when the phase starts to reach
the minimal band gap of the SSET at $\varphi = \pm \pi$. At the
second band $|L_{J}| \gg L$ and thus it has little effect on the
resonant frequency. Experimentally, close to $P_c$, both
resonances appear simultaneously, in agreement with the SSET being
at the second band for part of the time.

As seen in Fig.\ \ref{fig:gain}, the gain grows roughly like $(E_J
/ E_C)^{-1}$ at small $E_J / E_C$. $\partial | \Gamma | / \partial
l_{J}$ is a rather weak function of the parameters of the sample
or the tank circuit, however, small total inductance and matching
to $Z_0$ are favored. For low $E_J / E_C$, difficulties arise due
to noise sensitivity. Let us evaluate the ultimate performance of
the electrometer for $E_J / E_C = 0.15$ which we consider as still
usable with our present noise level.

The non-harmonic mode offers good possibilities to reach uncoupled
energy sensitivity of $\epsilon \approx \hbar$ in the present
configuration, using an Al SSET. Our simulations indicate that
$\partial | \Gamma | / \partial l_{J}$ is insensitive to $L_J$.
With the "optimal" Al SSET, $R_{SET} = 35$ k$\Omega$, $E_J = 0.45$
K and $L_{J} = 60$ nH, we have prediction $s_q \simeq 3 \times
10^{-7}$e$/\sqrt{\mathrm{Hz}}$. Since the minimal band gap between
the 1st and 2nd band in a SSET (in contrast to a Cooper-pair box
for instance) is insensitive to $E_J / E_C$, the decrease of
sensitivity due to mainly interband transitions amounts by roughly
the same factor $\eta \approx 4$ as in the present sample, thus we
expect to be feasible $\eta \times s_q \simeq 1 \times
10^{-6}$e$/\sqrt{\mathrm{Hz}}$, corresponding to $\epsilon
\lesssim \hbar$.

The best prospects to reach quantum-limited operation in the
plasma oscillation mode with $Q_{i} \leq 20$ are expected if
materials with higher $T_c$ than Al are used. Using rather
standard Nb-Al technique \cite{nbal}, with "effective" $\Delta =
\sqrt{\Delta_{\mathrm{Al}} \Delta_{\mathrm{Nb}}} \simeq 0.5$ mV,
$\eta \times s_q \simeq 1 \ldots 2 \times
10^{-6}$e$/\sqrt{\mathrm{Hz}}$ if a SQUID preamplifier with $T_N =
0.3$ K, which offers an additional benefit of a lower back-action
and thus less stringent conditions for noise isolation, is used.
In the non-harmonic mode, equally good sensitivities are simulated
for Nb-Al as for the case of Al.

Finally, we note that the system may provide a model system of a
qubit plus an integrated oscillator which works also as a detector
\cite{zorinqubit}, with an inherent advantage of efficient
filtering of external noise outside the band of the resonator.

In conclusion, we have demonstrated the feasibility of the L-SET,
where we use gate-dependent Josephson inductance of a
superconducting single-electron transistor for radio-frequency
electrometry. We foresee excellent prospects to operation close to
the quantum limit.

Fruitful discussions with T. Heikkil\"a, G. Johansson, R. Lindell,
H. Sepp\"a, and J. Viljas are gratefully acknowledged. This work
was supported by the Academy of Finland and by the Large Scale
Installation Program ULTI-3 of the European Union.


\begin{thebibliography}{99}

\bibitem{nakamuraqb} Y. Nakamura, Yu. A. Pashkin, and J. S. Tsai, Nature \textbf{398}, 786
(1999).
\bibitem{vion} D. Vion \emph{et al.}, Science \textbf{296}, 886 (2002).
\bibitem{hanqb} Y. Yu, S. Han, X. Chu, S. Chu, and Z. Wang, Science \textbf{296}, 889
(2002).
\bibitem{martinisqb} J. M. Martinis, S. Nam, J. Aumentado, and C. Urbina,
Phys. Rev. Lett. \textbf{89}, 117901 (2002).
\bibitem{ds} M. H. Devoret, and R. J. Schoelkopf, Nature \textbf{406}, 1039 (2000).
\bibitem{rfset} R. J. Schoelkopf, P. Wahlgren, A. A. Kozhevnikov, P. Delsing, and D. E. Prober, Science \textbf{280}, 1238 (1998).
\bibitem{zorin} A. B. Zorin, Phys. Rev. Lett. \textbf{76}, 4408 (1996).
\bibitem{zorinexp} A. B. Zorin \emph{et al.}, J. of Supercond.
\textbf{12}, 747 (1999).
\bibitem{zorinrf} A. B. Zorin, Phys. Rev. Lett. \textbf{86}, 3388 (2001).
\bibitem{cottet} A. Cottet \emph{et al.}, in \emph{Macroscopic Quantum Coherence and Quantum
Computing}, edited by D. Averin, B. Ruggiero, and P. Silvestrini
(Kluwer Academic, New York 2001), p. 111.
\bibitem{lukens} D. J. Flees, S. Han, and J. E. Lukens, Phys. Rev. Lett. \textbf{78}, 4817
(1997).
\bibitem{brink} A. Maassen van den Brink, L. J. Geerligs and G. Sch\"{o}n, Phys. Rev. Lett. \textbf{67}, 3030 (1991).
\bibitem{erne} S. N. Erné, H.-D. Hahlbohm, and H. L\"{u}bbig, J. Appl. Phys. \textbf{47}, 5440 (1976).
\bibitem{tbb} M. A. Sillanp\"a\"a \emph{et al.}, to be published.
\bibitem{qnote} Note that although a decrease in $Q$-value causes a
shift in $f_p$, this effect amounts here only to 300 kHz.
\bibitem{sample} Earlier sample with $E_J / E_C \approx 9$ gave
$s_q \approx 0.03$ e$/\sqrt{\mathrm{Hz}}$ in the non-harmonic
mode, in agreement with expectation $s_q \approx 0.02$
e$/\sqrt{\mathrm{Hz}}$.
\bibitem{nbal} N. Kim, K. Hansen, J. Toppari, T. Suppula, and J.
Pekola, J. Vac. Sci. Technol. B \textbf{20}, 386 (2002).
\bibitem{zorinqubit} A. B. Zorin, Physica C, \textbf{368}, 284 (2002).

\end{thebibliography}
\end{document}